\documentclass[
	reprint,
	longbibliography,
	twocolumn,
	aps,
	prd,
	preprintnumbers,
	nofootinbib]{article}
	
\pdfoutput=1

\usepackage[margin=2cm]{geometry}
\usepackage{inputenc}                           
\usepackage{graphicx}
\usepackage{float}
\usepackage[colorlinks=true,linkcolor=blue,citecolor=blue]{hyperref}  
\usepackage[noblocks]{authblk}             
\usepackage[dvipsnames]{xcolor}
\usepackage{cite}

\usepackage[T1]{fontenc}

\usepackage{enumitem}
\usepackage{latexsym}
\usepackage{amsfonts}
\usepackage{amssymb}
\usepackage{amsmath,multirow,bigdelim}
\usepackage{slashed}
\usepackage{dcolumn}
\usepackage{verbatim}
\usepackage{multirow}
\usepackage{xspace}
\usepackage[normalem]{ulem}
\usepackage[bottom]{footmisc}  
\usepackage{soul}                      
\usepackage{flushend}               
\usepackage{lipsum}                  
\usepackage{hyperref}

\usepackage{etoolbox}
\patchcmd{\thebibliography}{\section*{\refname}}{}{}{}

\def\met{\slashed{E}_T}
\newcommand\gev{\,\mathrm{GeV}}
\newcommand\tev{\,\mathrm{TeV}}
\newcommand\ifb{\, {\rm fb}^{-1}}

\newcommand{\beq}{\begin{equation}}
\newcommand{\eeq}{\end{equation}}
\newcommand{\bea}{\begin{eqnarray}}
\newcommand{\eea}{\end{eqnarray}}
\newcommand{\ba}{\begin{array}} 
\newcommand{\ea}{\end{array}}

\newcommand{\abs}[1]{\left|#1\right|}

\def\cs{\cos\theta_{\rm CS}}
\def\mtt{m_{T}^{\ell \nu \gamma}}
\def\mwa{m_{W\gamma}}

\definecolor{brown}{rgb}{.7,.35,.1}

\begin{document}

%
\title{ \large{ \bf The Radiation Valley and Exotic Resonances in $W\gamma$ Production at the LHC}}

\author[1,2,4]{\normalsize{Rodolfo M. Capdevilla}}
\author[3]{\normalsize{Roni Harnik}}
\author[4]{\normalsize{Adam Martin}}

\affil[1]{\normalsize{{\it Perimeter Institute for Theoretical Physics, Waterloo, Ontario N2L2Y5, Canada}}}
\affil[2]{\normalsize{{\it Department of Physics, University of Toronto, Toronto, Ontario M5S1A7, Canada}}}
\affil[3]{\normalsize{\it Theoretical Physics Department, Fermilab, Batavia, Illinois 60510, USA}}
\affil[4]{\normalsize{{\it Physics Department, University of Notre Dame, Notre Dame, Indiana 46556, USA}}}

\date{\vspace{-6ex}}

\twocolumn[
\begin{@twocolumnfalse}
\maketitle
\begin{abstract}
The tree-level partonic angular distribution of Standard Model $W\gamma$ production possesses a feature known as the Radiation Amplitude Zero (RAZ) where destructive interference causes the cross section to vanish. At the proton level the exact cancellation disappears, however, one can find a dip in the central region of the angular distributions, here called the Radiation Valley (RV). In this paper, we  show how the 
sensitivity for $W(\ell\nu)\gamma$ resonances can be significantly improved if one focuses on events in the RV region. Using this technique, we find that the LHC could probe a larger range of resonance masses, equivalent to increasing the luminosity by a factor of $2-3$ over conventional searches. The exact increase depends on the spin of the $W\gamma$ resonance and exactly how it couples to electroweak gauge bosons.
\end{abstract}
\vspace{2ex}
\end{@twocolumnfalse}]

\section{Introduction}\label{intro}

Electroweak diboson resonances ($X \to VV'$, where $V, V' \in \gamma, W^{\pm}, Z, H$) are a feature of many models ~\cite{Dorigo:2018cbl} of physics beyond the Standard Model (SM), including TeV-scale compositeness~\cite{Arkani-Hamed:2016kpz,Howe:2016mfq,Freitas:2010ht,Eichten:2007sx}, `quirkiness' \cite{Burdman:2008ek,Burdman:2014zta,Cheng:2018gvu}, heavy vector triplets \cite{Lizana:2013xla,Pappadopulo:2014qza}, and charged Higgses \cite{Logan:2018wtm,Song:2019aav}. While it may be tempting to view all diboson resonance searches as bump hunts, the SM background in certain electroweak diboson channels has some striking kinematic features that can be exploited to increase sensitivity. Specifically, the SM $W\gamma$ (and to some extent the $WZ$) channel exhibits a RAZ, an exact zero in the (tree level)  differential cross section $d\sigma/d\cos{\theta^*}$, where $\theta^{*}$ is the angle between the incoming parton and the outgoing photon in the center of mass frame \cite{Liu:2019vid}. For $W\gamma$ production, this zero occurs at $\cos\theta^{*}=\pm1/3$, where the negative (positive) sign corresponds to $W^+$ ($W^-$) production.  Kinematic regions where the SM background is suppressed are natural places to hone in on and search for new physics, as we explore in this paper. In particular, we investigate how to utilize the RAZ in $W\gamma$ to improve resonance searches by focusing the searches at the dip of the angular distributions. 

The RAZ phenomenon was discovered long ago by Brown, Mikaelian, Sahdev, and Samuel  \cite{Mikaelian:1979nr,Brown:1979ux} and it was soon after explained by Brodsky et al. \cite{Brodsky:1982sh,Brown:1982xx} as the quantum version of the classical result that there is no dipole radiation in the scattering of particles with the same $e/m$ ratio. The classical relation $Q_{1}/m_{1}=Q_{2}/m_{2}$ turns, at the quantum level, into $Q_1/(p_{1}\centerdot q)=Q_{2}/(p_{2}\centerdot q)$ \cite{Samuel:1983eg}, where $q_{1,2}$ are the charges of the colliding particles, $m_{1,2}$ their masses, $p_{1,2}$ their four-momentum, and $q$ is the four momentum of the outgoing photon. This formula defines the kinematic condition for which the amplitude of $W\gamma$ production is exactly zero.

\begin{figure}[t]
\begin{center}
	\includegraphics[width=.52\linewidth]{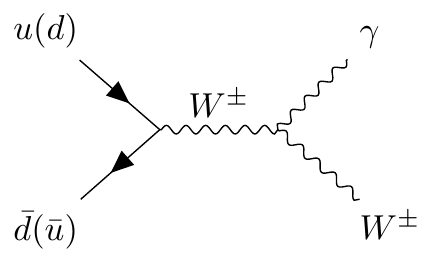} \\ \includegraphics[width=.52\linewidth]{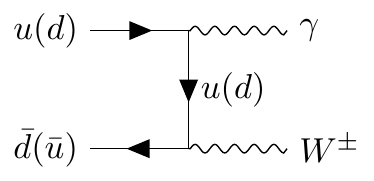}\includegraphics[width=.52\linewidth]{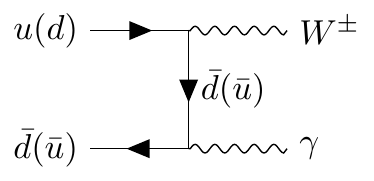}	
\end{center}
\caption{\label{diagrams}{\small Tree-level contributions to $W\gamma$ production ($s$, $t$, and $u$-channel respectively). For leptonically decaying $W$ one should add an extra diagram where the photon is emitted by the lepton (FSR).}}
\end{figure}

Diagrammatically, tree level, $W\gamma$ production occurs through the processes in Fig.~\ref{diagrams}. The RAZ exists due to a cancellation between these diagrams. Any new physics (NP) signal can modify the shape of the angular distribution by spoiling such cancellation. If the NP effects come from higher dimensional operators or anomalous triple gauge couplings the NP amplitude can interfere with the SM amplitudes and reshape the angular distribution in nontrivial ways. If the NP effects come from a resonance, the interference effects are subdominant, but the contribution from NP can populate the dip in the angular distributions where the SM contribution is zero. In this paper we will focus on the effects on the RAZ due to scalar and vector resonances in the context of a~few models.

In going from parton level to a realistic hadronic collision, the RAZ gets partially washed out by a number of effects \cite{Stroughair:1984pj,Baur:1988qt,Cortes:1985yi,Valenzuela:1985dp,Laursen:1983kw,Laursen:1982iv,Smith:1989xz,Ohnemus:1992jn,Baur:1989gk}. The largest contaminants are photon final state radiation (FSR), next-to-leading order corrections (NLO), and the reconstruction of the partonic center of mass frame. However, even including these washout effects, a clear dip in the angular distribution remains \cite{Baur:1993ir,Baur:1994sa}, as we will discuss below.

The layout of the rest of this paper is as follows: In section~\ref{RAZ} we present the parton RAZ and its proton version, the RV. We present the shape of the angular distributions for the background and a benchmark signal from a scalar resonance. This analysis helps us visualize the power of the RV as a kinematic cut that improves signal significance. Section~\ref{models} introduces a few models that include $W\gamma$ resonances. Section~\ref{results} presents our results, Brazilian flag-like exclusion regions for the signal cross section as a function of the resonance mass for different models. We finally present our conclusions in section~\ref{conclude}.

\section{The Radiation Amplitude Zero}\label{RAZ}

In this section we discuss how the RAZ manifests at parton and proton levels. For the latter we introduce a series of kinematic cuts that can be implemented to mitigate the washout effects due to NLO corrections and FSR. For this we will follow the ATLAS
 search for $W\gamma$ resonances in the leptonic channel in \cite{Aad:2014fha}. We then explore the behavior of the background and a benchmark signal model in different kinematic regions.

\begin{figure}[t]
	\includegraphics[width=0.95\linewidth]{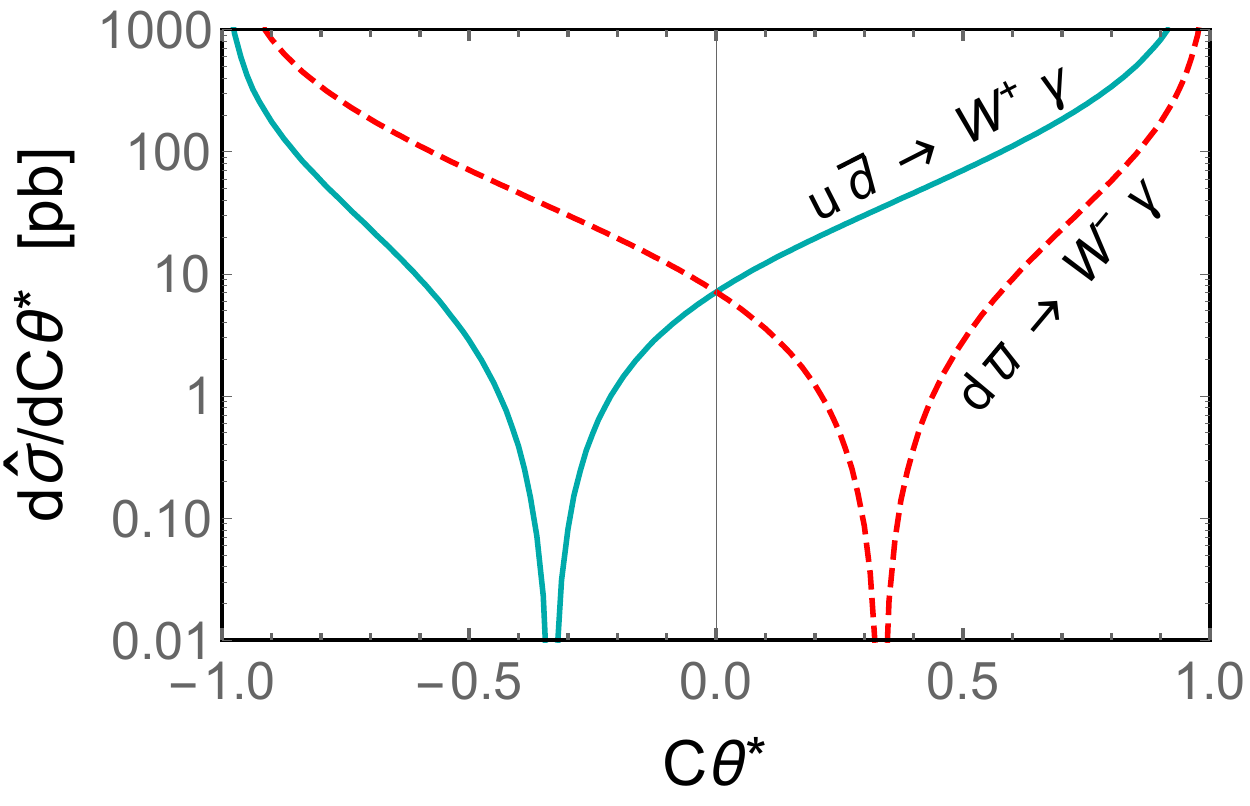}
\caption{\label{Cts}{\small Partonic angular distribution for $W^\pm\gamma$ production. The RAZ is located at $\cos\theta^*=\mp1/3$, where $\theta^*$ is the angle between the incoming quark and the photon in the CM frame.}}
\end{figure}

Starting at parton level, the tree-level partonic angular distribution of SM $W\gamma$ production shows an exact zero at $\cos\theta^*=\mp1/3$ for $W^\pm$ production. This zero in the cross section is the RAZ feature, shown in Fig.~\ref{Cts}.  As our goal is to use the RAZ to pick out regimes where the background is suppressed but the signal is not, we will also display the behavior of a benchmark NP model that decays isotropically to $W\gamma$. For now, we are only concerned with the shape of the signal, more details on the context for such a resonance will be shown in Sec.~\ref{sec:models}.

Moving from partons to protons, it is no longer possible to reconstruct the CM angle $\theta^*$ because we do not know the direction of the incoming quarks. However, there are several kinematic variables that encode some of the same information. One such variable is the rapidity difference between the photon and the lepton from the $W^+$ decay, $\Delta y = y_\gamma - y_\ell$,\footnote{Evidence of the RAZ have been found in measurements of $\Delta y$ at the Fermilab Tevatron \cite{Abazov:2008ad} and at CMS \cite{Chatrchyan:2011rr,Chatrchyan:2013fya}.} which has the benefit that it can be measured in the lab frame without worrying about reconstructing the four-momentum of the $W$.  A second variable is the Collins-Soper (CS) angle \cite{Collins:1977iv} , defined as \cite{Aad:2014wca,Khachatryan:2016yte}
\beq
\cs = \frac{Q_{z}}{|Q_{z}|} \frac{2(p_{1}^{+}p_{2}^{-}-p_{1}^{-}p_{2}^{+})}{|Q|\sqrt{Q^{2}+Q_{T}^{2}}}~,
\label{CS}
\eeq
where $Q$ is the net momentum of the $W\gamma$ system with $Q_z$ ($Q_T$) the longitudinal (transverse) piece, and $p_{i}^{\pm} = \left(p_{i}^{0}\pm p_{i}^{z}\right)/\sqrt{2}$, defining $p_{1}$ ($p_{2}$) as the momentum of the photon ($W$ boson). Determining the CS angle requires reconstructing the four-momentum of the leptonic $W$ boson. As there is only one source of missing energy in $W^+(\ell\nu)\,\gamma$ events, this reconstruction is possible (up to a two-fold ambiguity) and requires {\it i)} to assign all of the missing energy in the event to the neutrino, and {\it ii)} to assume that the $W$ boson is on-shell \cite{Panico:2017frx,Chiang:2011kq,Han:2009em,Novak:2018}.
\begin{figure*}[t!]
\begin{center}
\includegraphics[width=0.48\linewidth]{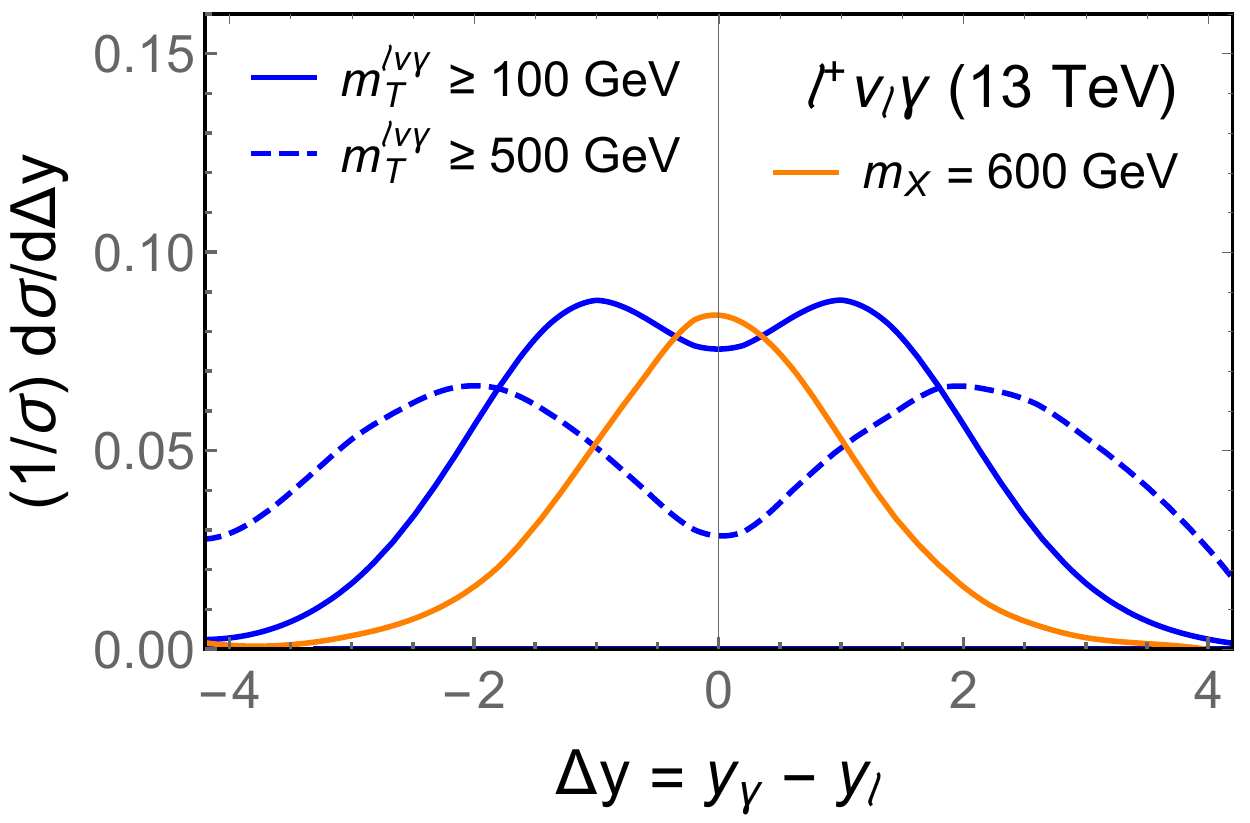}~~\includegraphics[width=0.48\linewidth]{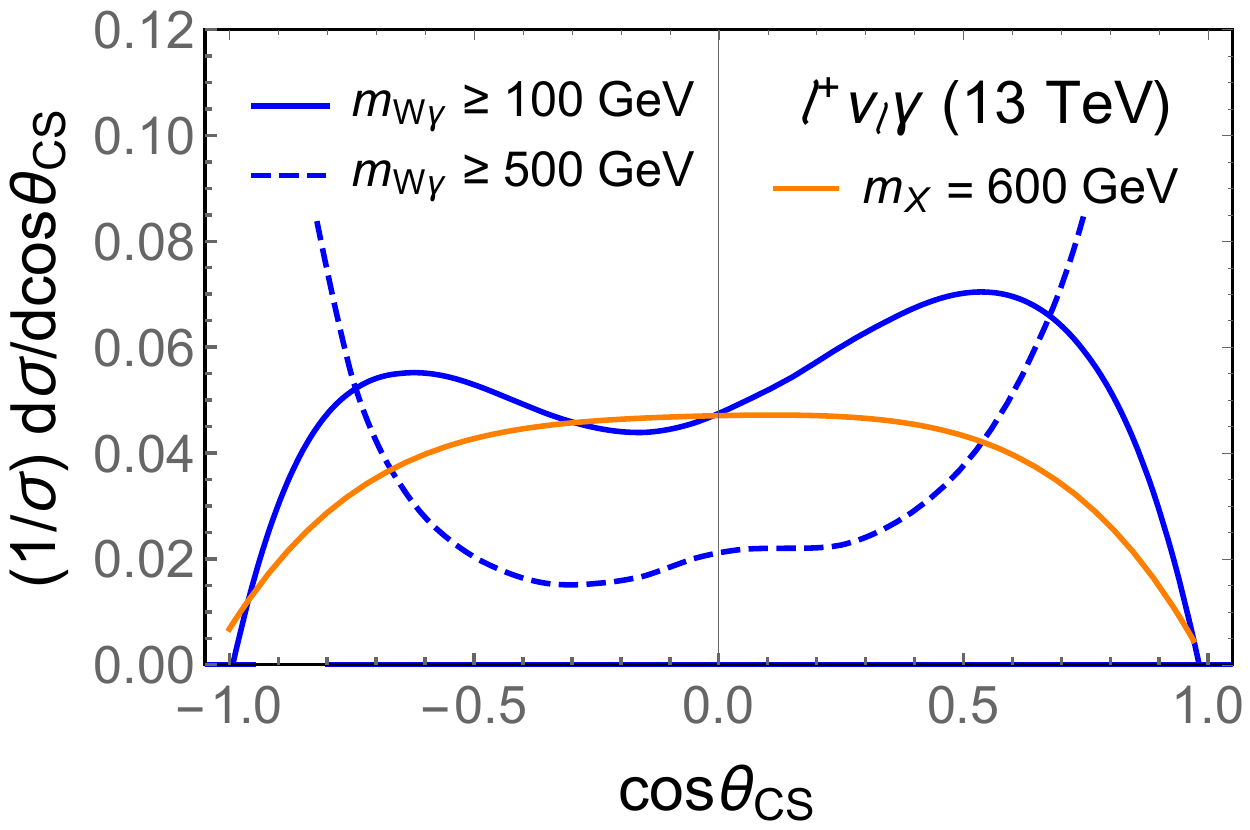}
\end{center}
\caption{\label{Proton}{\small Proton level distributions of $W^+(\ell^+\nu)\gamma$ production. The panels show the background and signal shape of two distributions: the difference of the photon and lepton rapidities $\Delta y$ (left), and the Collins-Soper angle $\cos{\theta}_{CS}$ (right). Solid lines show the distributions including the cuts in Eqs. \ref{eq:cuts}, \ref{eq:veto}, \ref{eq:FSR} only, while the dashed lines show the distributions after an extra cut in the cluster transverse mass $\mtt$ (Eq.~\eqref{eq:mt}) (left), or the invariant mass $m_{W\gamma}$ (right). The orange line represents a benchmark signal from the scalar triplet model introduced below, with a mass of $m_X = 600$ GeV.  All distributions are area normalized. The total cross section of the dashed $\Delta y$ ($\cs$) distribution corresponds to $\sigma=12$ fb ($\sigma=19$ fb).}}
\end{figure*}
 
To study how these variables capture the RAZ and how the $\Delta y$, $\theta_{CS}$ background and benchmark signal distributions vary across different kinematic regimes, we turn to Monte Carlo (MC).
For the background, distributions were calculated at NLO including FSR effects using {\tt MCFM 8.0}~\cite{Campbell:1999ah,Campbell:2011bn,Campbell:2015qma}. For the signal, we use the scalar triplet model (discussed in Sec.~\ref{sec:models}), having implemented it in FeynRules \cite{Degrande:2011ua} generating (LO) events using MadGraph5 \cite{1405.0301}. Our basic set of kinematic cuts is:
\begin{align}
p_{T}^{\gamma}   &   \ge50\gev, \quad \abs{\eta_{\gamma}}\le2.5,   \notag               \\
p_{T}^{\ell}           &   \ge30\gev, \quad \abs{\eta_{\ell}}\le2.5,            \label{eq:cuts}  \\
\met & \ge30\gev \notag.
\end{align}

These cuts must be supplemented by additional restrictions to mitigate the washout of the RAZ from higher order QCD and QED effects. Specifically, the impact on the RAZ from NLO effects is minimized by imposing a jet veto on the events \cite{Baur:1993ir,Baur:1994sa} whereas the impact due to the FSR contribution is minimized with cuts on the angle between the electron and the photon, $\Delta R_{\ell\gamma}$, and the photon isolation cone, $R_0$ \cite{Frixione:1998jh}:
\begin{align}
\text{jet\, veto}:&\quad p_T^j \ge 30\gev, \quad \abs{\eta_j}\le 2.8, \label{eq:veto}\\
\text{FSR\, veto}:&\quad \Delta R_{\ell\gamma}  \ge0.7,  \quad   R_0  \ge0.4.\, \label{eq:FSR}
\end{align}

Finally, we would like some way to evaluate how the signal and background shapes vary with the overall energy of the event. This can be done in two ways, depending on if leptonic $W$ reconstruction is possible or not: 1. We can use the invariant mass of the events $\sqrt{\hat s} = m_{W\gamma}$, or 2. We can use the transverse cluster mass $\mtt$ defined as~\cite{Aad:2014fha}: 
\begin{multline}
\left(\mtt \right)^2 = \left( \sqrt{m_{\ell\gamma}^2 + |\mathbf{p}_T^\ell + \mathbf{p}_T^\gamma |^2} + \met \right)^2 - \\ - \left| \mathbf{p}_T^\ell + \mathbf{p}_T^\gamma + \mathbf{p}_T^\nu \right|^2,
\label{eq:mt}
\end{multline} 
as a proxy for $\sqrt{\hat s}$. Here, $m_{\ell\gamma}$ is the invariant mass of the $\ell\gamma$ pair. In our analysis, we will consider both options.

The $\Delta y$ and $\cos{\theta_{CS}}$ distributions for $pp \to W^+(\ell^+\nu)\gamma$ are shown in Fig.~\ref{Proton}, along with the isotropic benchmark signal. All distributions are shown imposing the cuts in Eqs. \ref{eq:cuts}, \ref{eq:veto} and \ref{eq:FSR}. To explore how the distributions vary with the overall energy of the partonic collision, we consider two different values of $\mwa$, $\mtt \ge 100\,\gev$ (solid lines) and $\mwa,$ $\mtt \ge 500\,\gev$ (dashed lines). The benchmark signal assumes $m_X = 600\,\gev$, where $X$ is the $W\gamma$ resonance. As the signal $\mtt$  distribution is peaked near $m_X$, the $\mtt$ cut has little impact (provided its not too close to $m_X$), so we only show the signal curve for the higher cut. The distributions are area normalized so we can focus on the difference in shape.

The distributions in Fig.~\ref{Proton} show the following qualitative features:

\begin{itemize}
\item For $\Delta y$ (left), the SM distribution shows a dip in the central region~\cite{Barze:2014zba} related to the RAZ. For large values of the transverse mass, the depth of the central dip is bigger. On the other hand, the signal distribution is peaked at the central region.

\item For $\cs$ (right), the SM distribution has a minimum at the center-left angular region $\cos\theta_{CS} \in [-0.4,0.1]$. This feature is a clear indicative of the underlying RAZ. On the other hand, the signal populates the angular distribution in a nearly isotropic way in the central region.
\end{itemize}

We refer to the dips in the $\Delta y$ and $\cs$ distributions as the Radiation Valley (RV) feature. To be more quantitative about the RV, we calculate the efficiency of the RV cut, defined as the ratio
\beq
\epsilon_{V} = \frac{{\rm number~of~events~inside~the~valley}}{{\rm total~number~of~events}},
\label{rv}
\eeq
which can clearly be optimized depending on the scenario. For example, for $\mtt$ ($\mwa$) in the interval $[0.5, 0.7]$ TeV, using the RV cut $\Delta y \in [-1.2, 1.2]$ ($\cos\theta_{CS} \in [-0.7,0.5]$) the background has an efficiency of $\epsilon^{(b)}_{V} = 0.25$ ($0.28$) whereas the signal from the scalar benchmark model in Fig.~\ref{Proton} has an efficiency of $\epsilon^{(s)}_{V} = 0.72$ ($0.72$). If we quantify the significance of a signal by $S/\sqrt{B}$, where $S$ is the number of signal events and $B$ is the number of background events, the impact of a new cut changes the significance by $\epsilon^{(s)}/\sqrt{\epsilon^{(b)}}$. Plugging in the RV cut numbers for this example point, we find an increase in significance of $44\%$ ($36\%$).

Having seen how effective the RV cut can be at reducing the SM $W\gamma$ background, our strategy is simple: we will supplement current $W\gamma$ resonance search cuts with our RV cut, with the valley optimized for a given resonance spin and mass. 

\begin{figure}[t!]
	\includegraphics[width=1.0\linewidth]{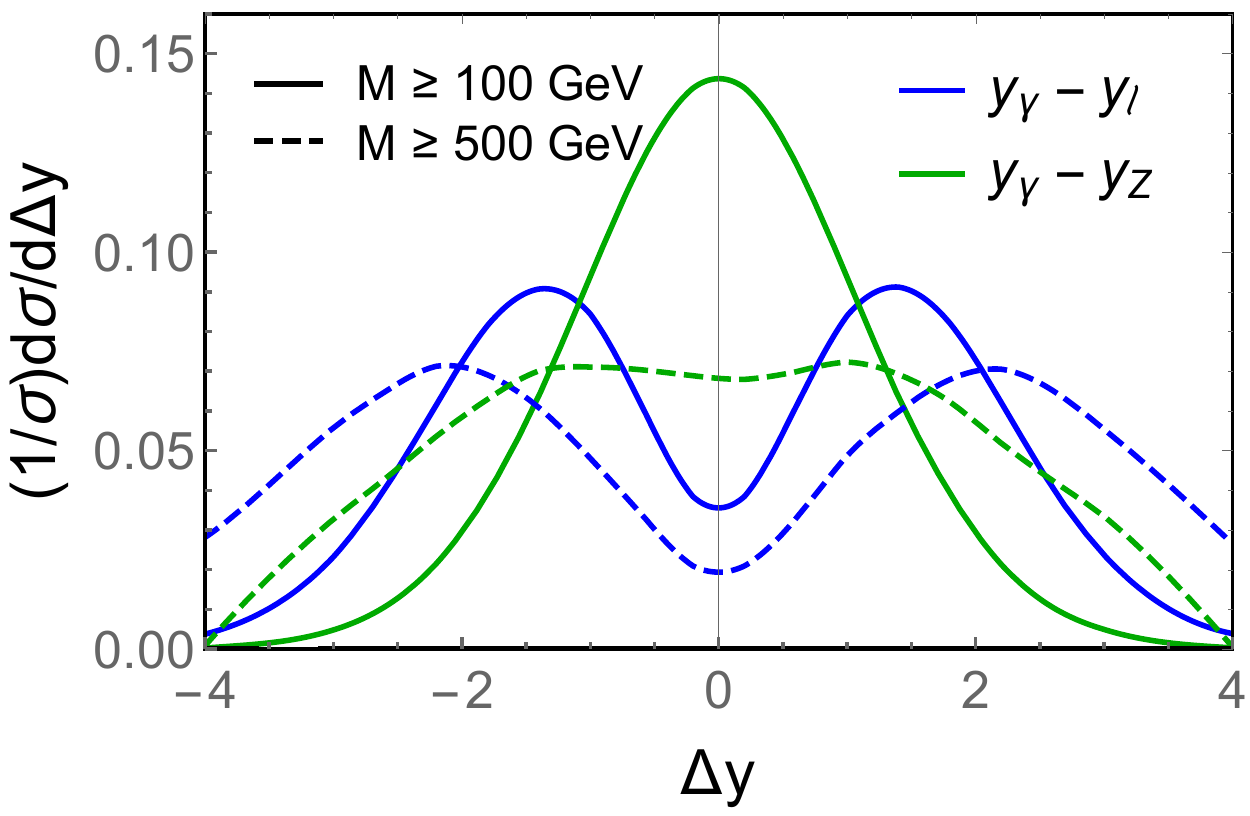}
\caption{\label{Zgamma}{\small Comparing difference in rapidities distributions from $W(\ell^+\nu)\gamma$ and $Z(\ell^+\ell^-)\gamma$ production (SM background at LO). Solid lines include the basic cuts in Eqs. \ref{eq:cuts}, \ref{eq:FSR}, while dashed lines include an extra cut in the transverse (invariant) mass $\mtt$ ($m_{Z\gamma}$) in $W\gamma$ ($Z\gamma$) indicated by the letter $M$.}}
\end{figure}

Before going forward with this strategy, it is important to differentiate between SM processes with a RAZ from generic SM diboson processes. As can be seen from Fig.~\ref{Proton}, the SM $W\gamma$ distributions are pushed to large values of $|\Delta y|$, $\cos{\theta_{CS}}$ as $\hat s$ increases. Some of this can be attributed to $t/u$-channel contributions to $W\gamma$, which are enhanced at $\theta \to 0,\pi$, or, said differently, are less suppressed at $\theta \to 0,\pi$ when $\hat s$ is large. However, part of the dip is due to the cancellation between the $s$-channel contribution with the $t/u$-channels. To highlight the impact of the cancellation on the $\Delta y$, $\cos{\theta_{CS}}$ distributions, in Fig.~\ref{Zgamma} we compare the  $\Delta y$ distributions (area normalized) of (LO) $W\gamma$ and $Z\gamma$, a diboson process with $t/u$-channel pieces only. For both processes, we impose the cuts in Eqs. \ref{eq:cuts}, \ref{eq:veto}, and \ref{eq:FSR}. For $Z\gamma$ we apply the lepton cuts to both leptons in the $Z$ decay and the invariant mass of the leptons is restricted to $65 < m_{\ell\ell} ({\rm GeV}) < 115$. We can see how the $Z\gamma$ process starts to develop a central dip at high enough energies due to its $t/u$ channel nature, but the dip is clearly deeper for $W\gamma$. More quantitatively, we can compare the effect on $Z\gamma$ caused by the RV cut that we used for $W\gamma$ right below Eq.~\eqref{rv}. For the same cut values we used for the $W\gamma$ scenario --  $\mtt$ in the interval $[0.5, 0.7]$ TeV and RV cut $\Delta y \in [-1.2, 1.2]$, $Z\gamma$ has an efficiency of $\epsilon^{(b)}_{V} = 0.42$ while a hypothetical scalar $Z\gamma$ resonance has an efficiency of $\epsilon^{(s)}_{V} = 0.56$. These values represent a {\em decrease} of $14\%$ in the significance\footnote{Optimizing the cut values for $Z\gamma$ rather than using the same ones we used for $W\gamma$ does not significantly improve the situation.}. In this sense, the RAZ provides a distinct advantage for $W\gamma$ over other channels in searches for exotic diboson resonances.

Two additional comments worth mentioning: First, the results in this and subsequent sections focus on $W^+$ production for simplicity. One can, in principle, apply the same analysis in $W^-$ production. Second, the most recent analysis by ATLAS with $W\gamma$ looks at hadronic decays of the $W$ boson~\cite{Aaboud:2018fgi}. The analysis utilizes substructure techniques for highly boosted fat jets to disentangle $W\gamma$ from processes like $Z\gamma$, $H\gamma$ and large QCD backgrounds, a complication that we want to avoid by limiting our analysis on the leptonic channel. Focusing on leptonic $W$ also allows us to distinguish $W+$ from $W^-$, which is beneficial since the location of the RV depends on the $W$ charge.

Now that we have identified the RV cut strategy and verified its connection to the underlying RAZ in the SM background, we are ready to introduce some models with $W\gamma$ resonances.

\section{Models for $W\gamma$ Resonances}\label{models}
\label{sec:models}

The first model we will analyze with the RV cut are scalar quirks in Folded Supersymmetry \cite{Burdman:2008ek,Burdman:2014zta}. 
Supersymmetry is an attractive way to ensure the cancelation the quadratic divergences of the Higgs mass parameter. 
In the Minimal Supersymmetric Standard Model each superpartner is charged under the same symmetries as its corresponding SM particle. This is dictated by the symmetry structure of the model; supersymmetry commutes with gauge symmetries. By contrast, in Folded Supersymmetry the quark superpartners are not charged under the SM color $SU(3)_{c}$, but rather under a mirror (or dark) color $SU(3)_{c'}$ (the superpartners remain charged under the SM electroweak interactions). Supersymmetry is broken in the ultra-violet (UV) in a way that respects a $Z_2$ symmetry  that relates both $SU(3)_{c} \leftrightarrow SU(3)_{c'}$. This still guarantees the cancelation of the quadratic divergences in the IR at one-loop~\cite{Burdman:2006tz}. However, folded superpartners are not colored under SM QCD and thus produced at electroweak rates.

Due to the $Z_2$, the confining scales of $SU(3)_c$ and $SU(3)_{c'}$ forces are similar. The direct bounds on charged particles imply that the lightest particles charged under $SU(3)_{c'}$ are much heavier than the hidden color confining scale. Particles that obey this criterion, being charged under a confining force without light matter, are dubbed `Quirks'~\cite{Kang:2008ea,Harnik:2008ax,Cai:2008au,Kribs:2009fy,Craig:2015pha,Craig:2016kue,Knapen:2017kly}, and in our model `squirks' (they are sometimes called F-squirks, but we will drop the F for ease of pronunciation).
The hallmark of quirky phenomenology is that upon production, quirks will form confined, but highly excited bound states among themselves. 

We will use $\tilde{q}$ to represent the squirk and $\left\langle \tilde{q}\tilde{q}^{*}\right\rangle^{Q}$ to represent the quirky bound state, or quirkonium, with charge $Q$.  Squirks production at the LHC begins with electroweak pair production. As the squirk electroweak charges are governed by supersymmetry, the only free parameter this process depends on is the squirk mass. The produced squirks then bind to each other via $SU(3)_{c'}$  forming a quirky bound state. The bound state can be either neutral or electrically charged: 
$$q\bar{q}\to\gamma^*/Z\to\tilde{q}\tilde{q}^{*}\to({\rm confining})\to\left\langle \tilde{q}\tilde{q}^{*}\right\rangle ^{0},$$ $$q'\bar{q}\to W^\pm\to\tilde{q}'\tilde{q}^{*}\to({\rm confining})\to\left\langle \tilde{q}'\tilde{q}^{*}\right\rangle^{\pm}.$$ 
Once formed, there are several processes that a bound state can undergo. 
First, as the quirkonium is formed in a highly excited state, it can relax down to the 1S state by emitting soft photons and dark glueballs \cite{Burdman:2008ek}. Second, charged quirkonium can decay to neutral quirkonium via the beta decay $\tilde{q}' \to W(\ell\nu) \tilde{q}$, $$q'\bar{q}\to W^\pm\to\tilde{q}'\tilde{q}^{*}\to \ell^\pm \nu \left\langle \tilde{q}\tilde{q}^{*}\right\rangle^0.$$ The rate for beta decay depends crucially on the mass splitting between $\tilde{q}'$ and $\tilde{q}$. Finally, quirkonium can decay back to SM particles. Assuming the bound states  relax down to their ground state before decaying, they behave as scalar resonances. Neutral quirkonium decay primarily to dark glueballs, whereas charged ones decay to $W\gamma, WZ$ and SM fermions, with the exact fraction depending on the relative velocity of the squirk constituents when they annihilate~\cite{Burdman:2008ek}. 

In order for squirks to act as $W\gamma$ resonances, we therefore need beta decay to be suppressed and the $W\gamma$ decay mode to dominate. Following the model in Ref.~\cite{Burdman:2008ek}, these conditions are satisfied for first and second generation folded squirks (which all have similar mass). For these states, taking the relative velocity to be small, the $W\gamma$ branching fraction is $85\%$. Third generation squirks do not contribute as $W\gamma$ resonances because the mass difference between the third generation squirks is large and $\langle\tilde t'\,\tilde b^*\rangle^+$, etc. promptly undergoes beta decay.

The second scenario we explore is more phenomenological -- composite $W\gamma$ resonances from some new TeV-scale strong dynamics. This scenario can be broken down further according to the spin of the composite, either scalar/pseudoscalar composites or vector composites, which can be thought of as pions or rho mesons of the new strong dynamics \cite{Freitas:2010ht,Howe:2016mfq,Eichten:2007sx,Dorigo:2018cbl,Arkani-Hamed:2016kpz,Bai:2010mn,Kilic:2008pm,Kilic:2009mi,Kilic:2010et}. Due to electromagnetic gauge invariance these composite states can decay to gauge bosons only through higher dimensional operators. In the following we provide two toy models that will allow us to quantify the power of our RV cutting technique. The UV completion of these models lies outside the scope of this paper.

\begin{itemize}
\item Scalar Triplets: Scalar composite states can have anomaly-induced interactions of the form $(1/\Lambda) \phi^{a}W_{\mu\nu}^{a}\tilde{B}^{\mu\nu}$, where $\Lambda$ is related to the scale at which the UV theory confines. Here $\phi^a$ is a pseudo scalar $SU(2)_L$ triplet. We consider this interaction along with a coupling to matter, $y_m$,
\beq
\mathcal{L}\ni-\frac{y_{m}}{\Lambda}\left(iQ^{\dagger}t^{a}Hd^{c\dagger}\phi^{a}+{\rm h.c.}\right)+\frac{1}{\Lambda}\phi^{a}W_{\mu\nu}^{a}\tilde{B}^{\mu\nu}.
\label{scalar}
\eeq
Note that we have chosen the triplet $\phi^a$ to be a pseudoscalar so that its neutral component does not mix with the Higgs (assuming CP conservation)~\cite{Howe:2016mfq}.

\item Vector Triplets: Vector composite states can couple directly to matter through a renormalizable interaction, and to gauge bosons through higher dimensional operators. We consider the following interactions for our analysis
\begin{multline}
\mathcal{L}\ni-g_{m}Q^{\dagger}t^{a}\bar{\sigma}^{\mu}QV_{\mu}^{a}+\frac{c_{W}}{\Lambda^{2}}V_{\mu}^{a\nu}W_{\nu}^{a\alpha}B_{\alpha}^{\mu} + \\ + \frac{c_{h}}{\Lambda^{2}}V_{\mu}^{a}B^{\mu\nu}H^{\dagger}t^{a}\overleftrightarrow{D}_{\nu}H,
\label{vector}
\end{multline}
where $g_m$ is the coupling to matter and the operator $c_W$ and $c_h$ provide two examples through which the heavy vector can decay to gauge bosons. The terms differ on how the $W$ boson appears; as $c_W$ contains a field strength it contains (predominantly) couplings between $V_\mu^a$ and transverse $W$, while $c_h$, which involves the Goldstone degree of freedom via $D_\nu H$, couples $V_\mu^a$ (predominantly) to longitudinal $W$. The vector $V_\mu^a$ can in principle mix with some of the vector bosons, with mixing typically $\mathcal{O}(m^2_W/m^2_{V})$ \cite{Dobrescu:2015jvn}. As we will focus on $m_V$ in the $0.5 - 1.5\, \tev$ range, this mixing is always small and we will ignore it. Had we coupled $V^a$ to leptons, there would be a strong constraint coming from resonant dilepton searches, hence we assume our vector resonance benchmark is leptophobic.

For simplicity, in our analysis we will include only one of these operators at a time. We define the {\it Vector Triplet $c_W$} model with the Lagrangian above choosing: $c_W=1, c_h=0$, and the {\it Vector Triplet $c_h$} model by choosing: $c_W=0, c_h=1$.
\end{itemize}

The diboson resonances $\phi^a$ and $V_\mu^a$ are produced through $q'\bar{q}$ collisions. This production is controlled by the matter couplings $y_m$ and $g_m$, respectively. 
The decays of the triplets are controlled by $y_m, g_m$, along with the interactions with SM gauge bosons which are proportional to the UV scale $\Lambda$.  We restrict our parameter space to regions where the resonances are narrow and the $W\gamma$ branching fraction is large enough so that this channel is more sensitive than channels like dijets. For the scalar model, this is not a strong limitation as all couplings are higher dimensional so the restriction $y_m <1$ already does the job. The vector models are more restricted in that sense because the decay $V^a \to q\bar q$ is not suppressed by $\Lambda$. As a result, $V^a$ will decay predominantly into dijets unless the coupling of $V_a$ to quarks $g_m$ is small enough. Finally, to make sure our effective resonance theories are under control, we will require that $\Lambda$ is at least 5 times bigger than the mass of the triplets.

Having laid out the models and the search strategy, we now move on to show the results.

\section{Results}\label{results}

We begin with a review of the current search for $W\gamma$ resonances performed by the LHC in the leptonic channel \cite{Aad:2014fha} and we then show how the results can be improved using our technique. The ATLAS collaboration searches for $W\gamma$ resonances in the leptonic channel using $20.3 \ifb$ of 8 TeV data~\cite{Aad:2014fha}. The 7 TeV analysis by CMS on $W\gamma$ final states was used to constrain anomalous triple-gauge couplings \cite{Chatrchyan:2013fya}, whereas more recent analysis focus on $Z\gamma$ production only \cite{Sirunyan:2017hsb}. We will follow these analysis and focus on the leptonic channel.

In the ATLAS analysis \cite{Aad:2014fha}, $\ell\nu\gamma$ candidate events are selected by requiring the following cuts similar to our Eq.~\eqref{eq:cuts}:  $p_T^\ell > 25$ GeV, $p_T^\gamma > 40$ GeV and $\met >$ 35 GeV, $|y_e|<2.47$, $|y_\mu |<2.4$, $|y_\gamma |< 2.37$ (excluding the calorimeter transition region $1.37 < |y_{\gamma,e} | < 1.52$). In addition, the transverse mass of the $\ell\met$ system is required to be greater than 40 GeV and the $\ell\gamma$ invariant mass needs to be outside the range $75-105$ GeV in order to suppress the background where one of the electrons from a $Z$ boson decay is misidentified as a photon. Finally, a selection requirement $\Delta R_{\ell\gamma}\ge 0.7$ is applied to suppress the FSR contributions (as in Eq.~\eqref{eq:FSR}). For events passing these cuts, ATLAS reconstructs the $W$ and hunts for bumps in the $\mtt$ distribution. As a template signal, they use the Low Scale Technicolor model in \cite{Eichten:2007sx} and set observed limits on cross section times branching ratio of about $1$ fb, ruling out resonance masses of up to $900$ GeV.
 
Starting with the ATLAS analysis, we add the RV cut. As discussed in Section \ref{RAZ}, the RV cut can be implemented on either $\Delta y$ or $\cos{\theta}_{CS}$, and the cut should be optimized for each mass bin -- quantified either with $m_{W\gamma}$ or the cluster transverse mass $\mtt$. We will explore all four combinations of RV angle and mass measurement, for each of the resonance signal models (squirks, Scalar triplet, Vector Triplet $c_W$, and Vector Triplet $c_h$).

\begin{figure}[t]
	\includegraphics[width=1.0\linewidth]{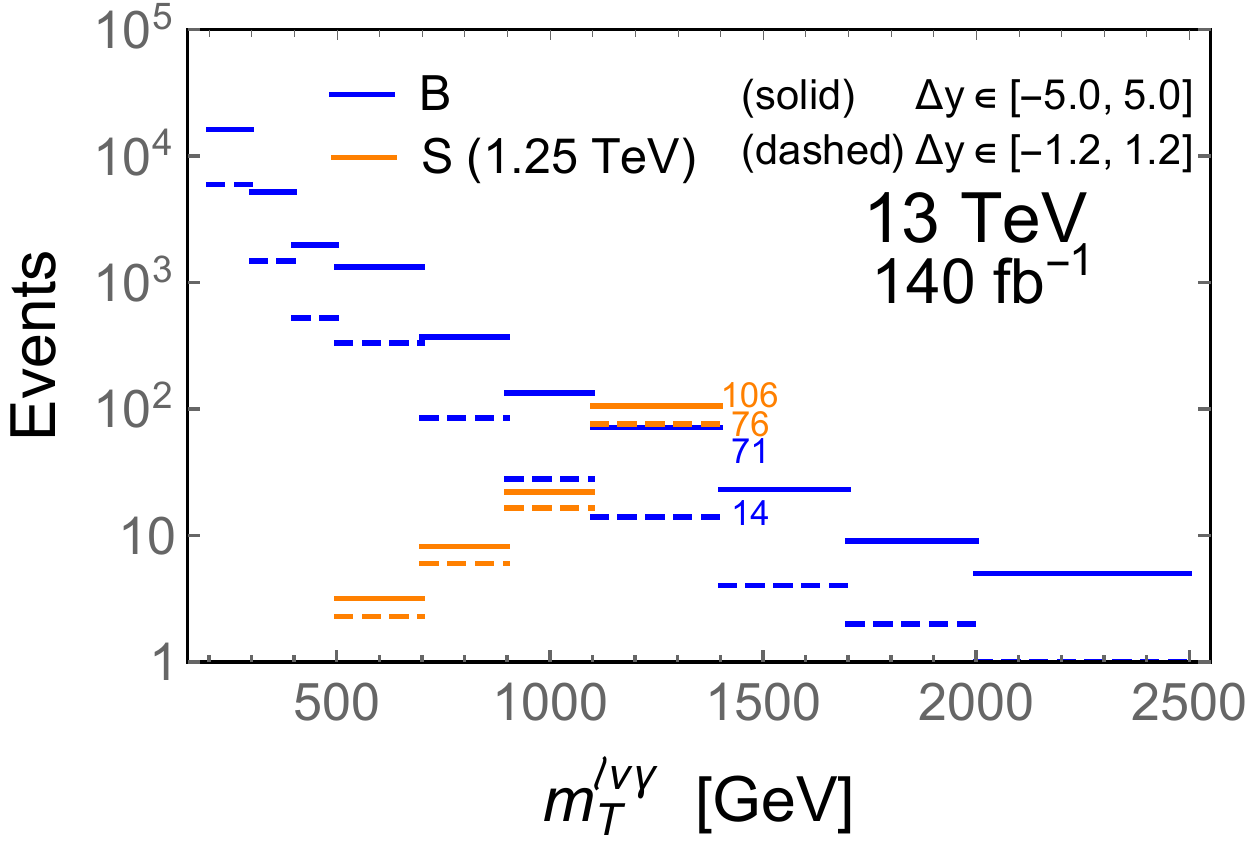}
\caption{\label{mTcut}{\small Transverse cluster mass distribution for $W^+(\ell^+\nu)\gamma$ production; background (blue) and a benchmark signal (orange) using the triplet scalar model introduced in Eq.~\eqref{scalar}. Solid lines include all events in the $\mtt$ bins, whereas dashed lines only include events for which $\Delta y \in \lbrack -1.2,1.2 \rbrack$. The small orange and blue numbers represent the number of events in the 1.25 TeV bin for signal (orange) and background (blue) without (solid) and with (dashed) the $\Delta y$ cut.}}
\end{figure}

To illustrate the impact of the RV cut, the pre-RV and post-RV $\mtt$ distributions for a $1.25\,\tev$ scalar triplet are compared in Fig.~\ref{mTcut}, with RV defined using $\Delta y$. The solid lines represent the background (blue) and signal (orange) $\mtt$ distributions including all values of $\Delta y \in \lbrack -5.0,5.0 \rbrack$, while the dashed lines impose the RV cut $\Delta y \in \lbrack -1.2,1.2 \rbrack$. The model parameters for this benchmark signal are $y_m=0.8$, and $\Lambda=10$ TeV, corresponding to a cross section times branching ratio (pre-RV) of $1$ fb. If we neglect any secondary backgrounds and estimate the significance using $S/\sqrt B$, the increase for this mass point is $60\%$. This example also shows the complications of $\mtt$, namely that the signal is not confined to a single bin. Depending on the resonance mass, roughly $20-30\%$ of events leak to lower $\mtt$.\footnote{This indicates that the bin size can be optimized for each signal  mass point. For example, increasing the size of the $1.25 \tev$ bin in Fig.~\ref{mTcut} would include more signal events as well as more background events, so there must be a bin size for which the significance is maximal. However, for simplicity, we will do our analysis using the binning showed in Fig.~\ref{mTcut} for all our signal models}

So far, we have only discussed the irreducible SM background coming from $W \gamma$. Reducible (fake) backgrounds, while subdominant overall, can play an important role as they may behave differently under the RV cut. One of the largest fake background comes from $W + \text{jet}$ where the jet fakes a photon ~\cite{Aad:2014fha,Chatrchyan:2013fya}. To incorporate this background in our analysis, we first estimate the jet $\to$ photon misidentification rate in ATLAS by comparing the NLO cross section in $W + \text{jets}$ (at 8 TeV, and after applying the analysis cuts) to the W + fake background quoted in Ref.\cite{Aad:2014fha}. This corresponds to a misidentification factor of $2\times10^{-4}$, which we conservatively assume to be constant in $p_T$ and $\eta$, and to carry over to analyses at 13 TeV. We then generate $W + \text{jet}$ at LO using {\tt MCFM 8.0}, treating the additional jet as a photon for the purposes of analysis cuts (including RV), and scaling the rate by the $2\times10^{-4}$ misidentification factor to determine the impact on the significance. (Following this procedure, we found that $W + \text{jet}$ represents a $8-10\,\%$ background). To estimate the effects of other reducible backgrounds and systematics we introduce by hand a systematic error comparable to the statistical error in each $\mtt$ and $\mwa$ bin. 

We implement the RV cutting procedure for all four models described in Sec.~\ref{sec:models} and throughout the mass range $0.3-2.5$ TeV. For each model, RV angle, mass variable ($m_{W\gamma}, \mtt$) and resonance mass we quantify the signal strength that is excluded by $2\sigma$ using Poisson statistics and $140\,\ifb$ of integrated luminosity.  Finally, we repeat the analysis for the HL-LHC, luminosity $3\, \text{ab}^{-1}$. In extrapolating to the HL-LHC, we assume that the efficiency of the signal and the background are unchanged, as is the behavior of the backgrounds.

\begin{figure*}[h!]
	\includegraphics[width=0.48\linewidth]{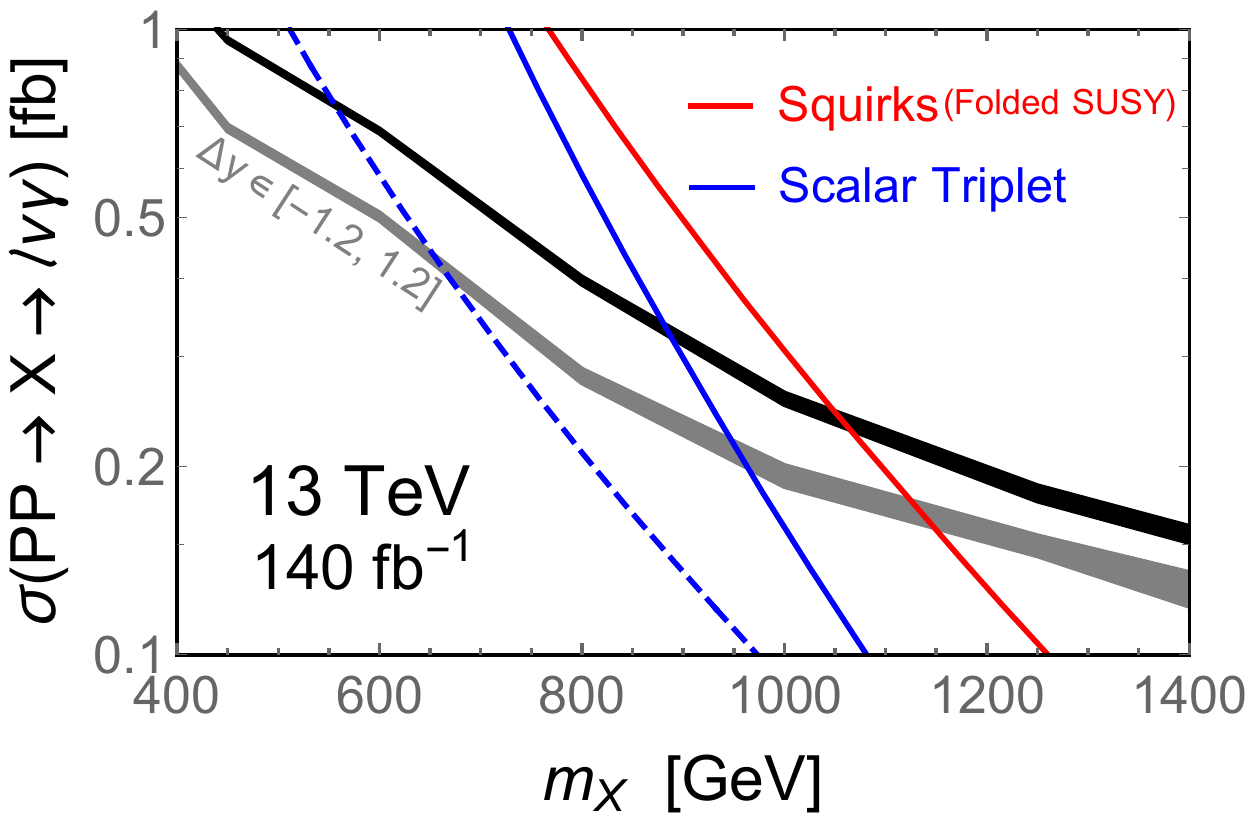}
	\includegraphics[width=0.505\linewidth]{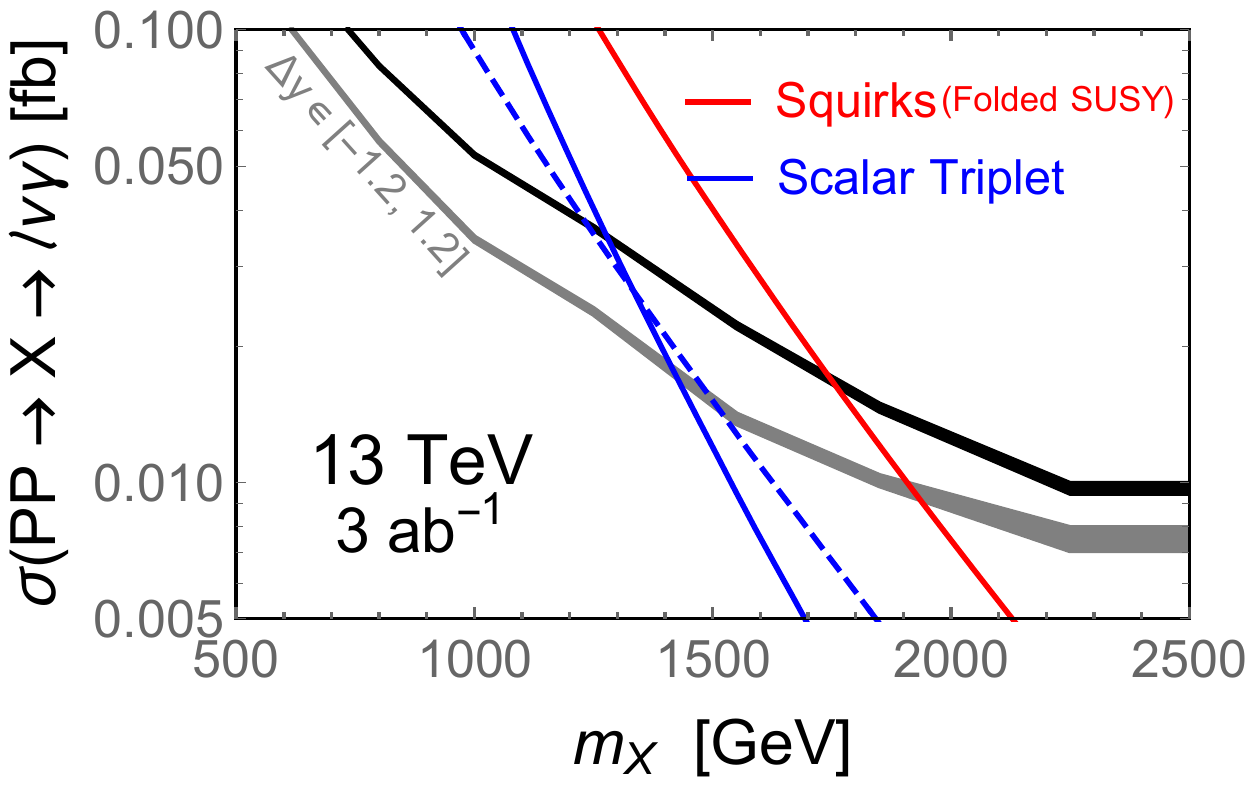}\\
	\includegraphics[width=0.48\linewidth]{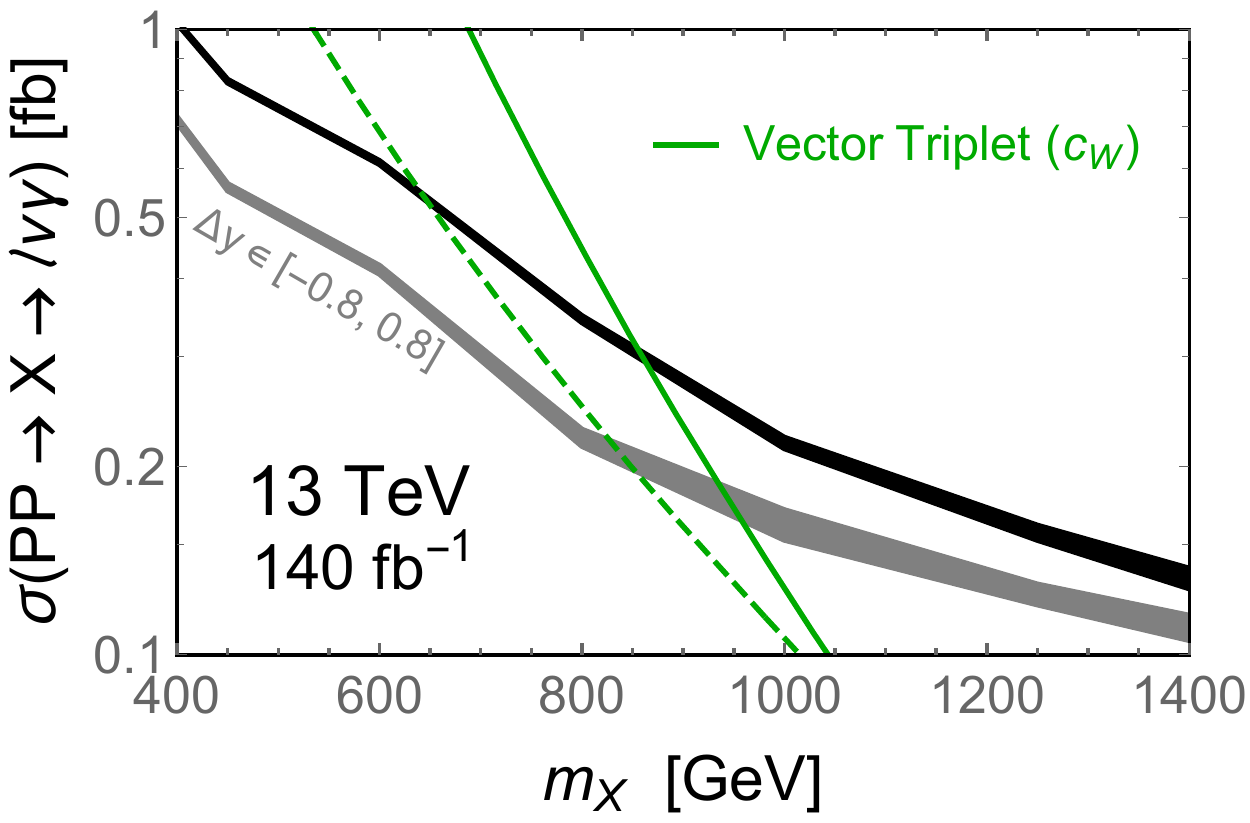}
	\includegraphics[width=0.505\linewidth]{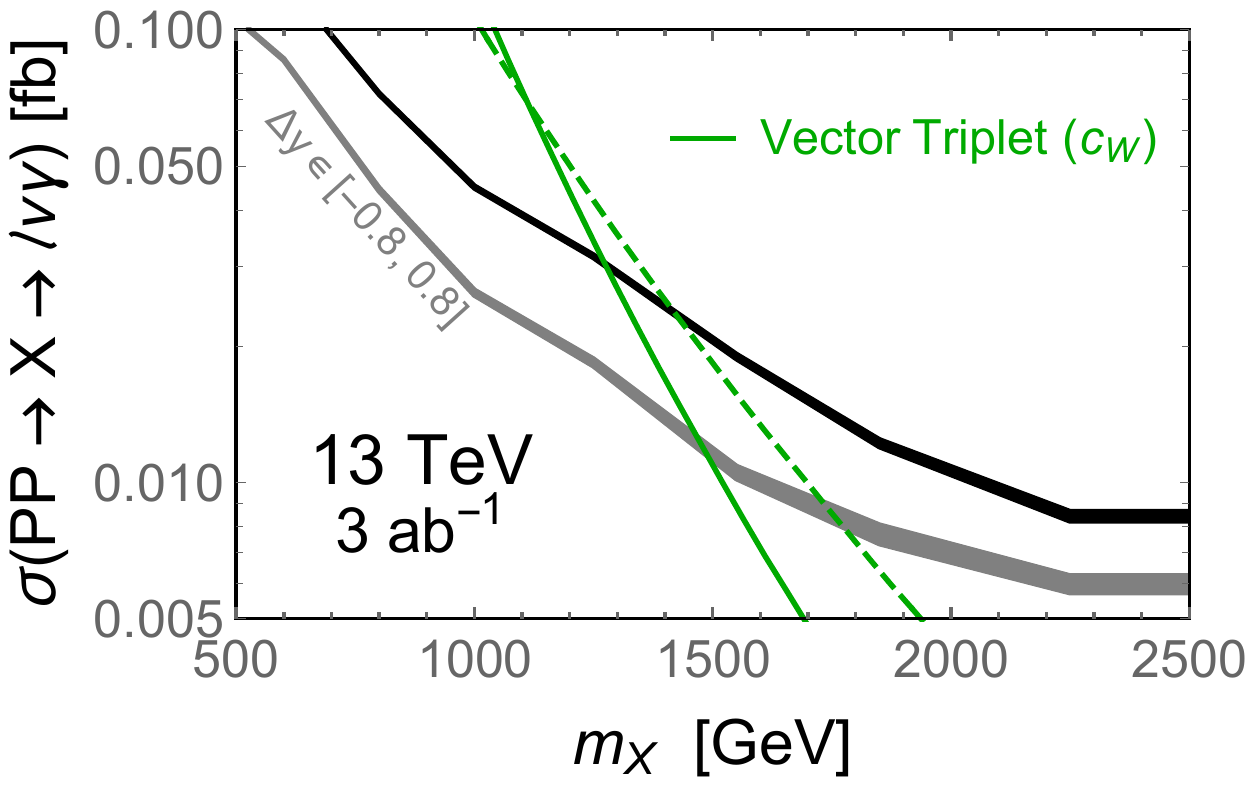}\\
	\includegraphics[width=0.48\linewidth]{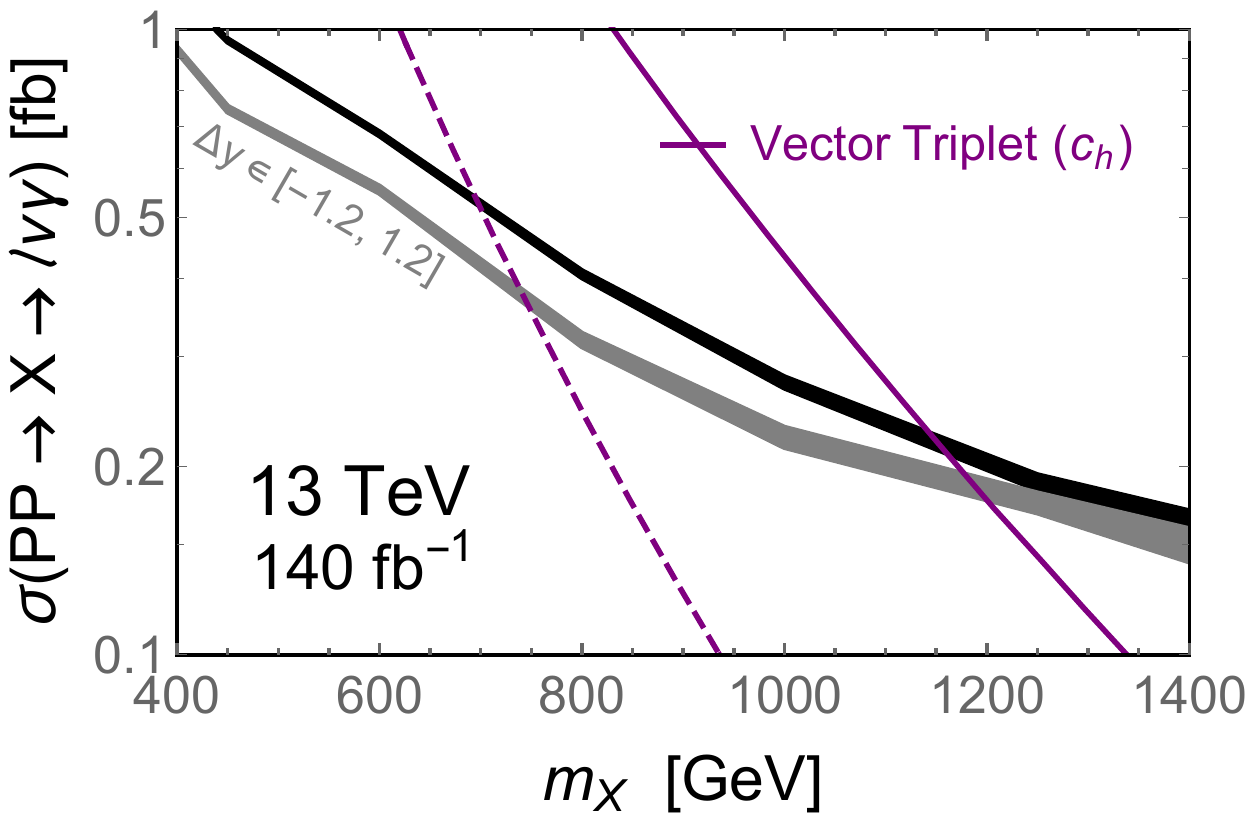}
	\includegraphics[width=0.505\linewidth]{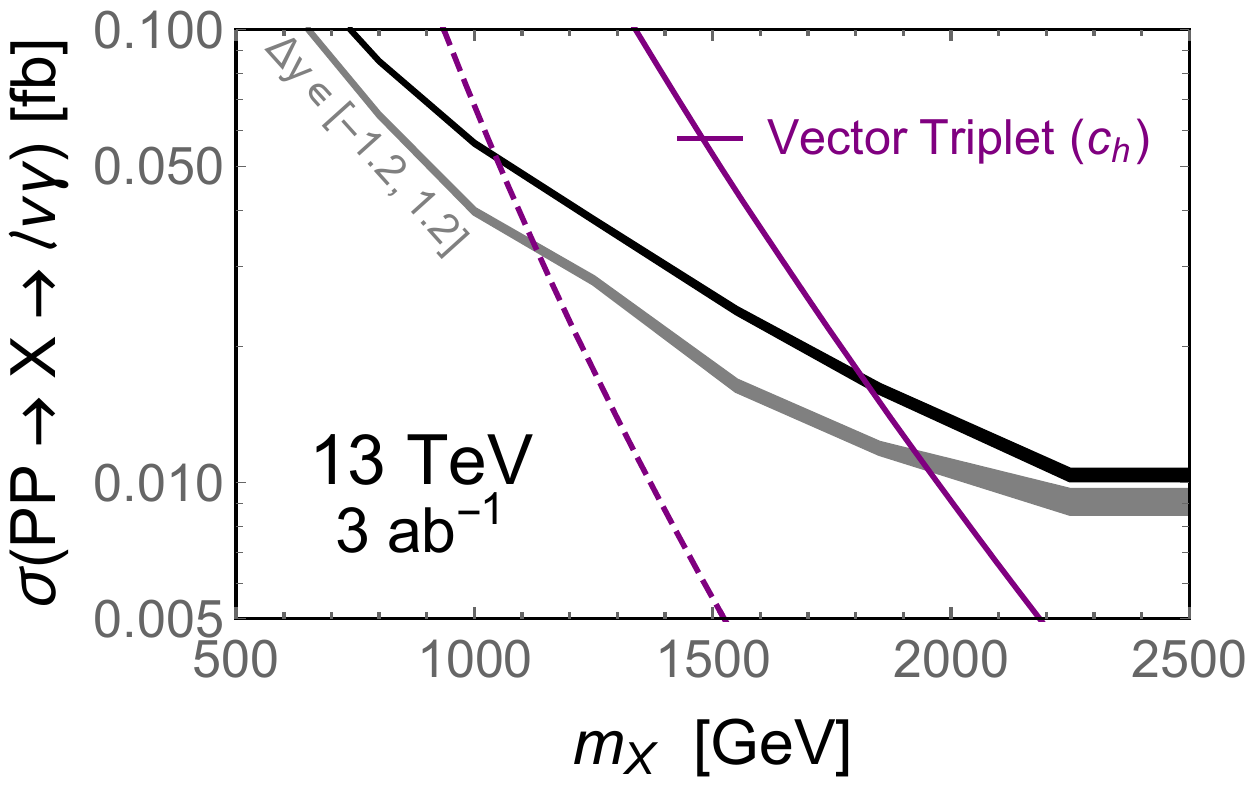}
\caption{\label{mass}{\small Sensitivity of $W\gamma$ resonances of mass $m_X$ at the $\sqrt s = 13\, \tev$ LHC assuming the current luminosity ($140\,\ifb$) (left) and high luminosity (right). Top: Squirks in Folded Supersymmetry including two generations of squarks (red), and the Scalar Triplet model fo Eq.~\eqref{scalar} using the benchmarks $\Lambda = 5\,m_X$, $y_m = 0.1$ (solid) and $\Lambda = 10$ TeV, $y_m = 0.15$ (dashed). Middle: Vector Triplet $c_W$ model with benchmarks $\Lambda = 5\,m_X$, $g_m = 0.03$ (solid) and $\Lambda = 10\,m_X$, $g_m = 0.06$ (dashed). Bottom: Vector Triplet $c_h$ model with benchmarks $\Lambda = 5\,m_X$, $g_m = 0.005$ (solid) and $\Lambda = 10\,m_X$, $g_m = 0.02$ (dashed). The black bands were calculated assuming the signal model shown in each panel and using the entire $\mtt$ distribution whereas the gray bands use $\mtt$ along with RV cuts on $\Delta y$ as shown in the gray label. The width of the bands come from considering a systematic error between once and twice as large as the statistical error in each $\mtt$ bin.}}
\end{figure*}

Putting everything together, we find that the optimal $W\gamma$ resonance mass sensitivity is always achieved imposing an RV cut on the rapidity difference $\Delta y$ in bins of the cluster transverse mass $\mtt$. The fact that $\mtt$ provides better results than $\mwa$ is a bit counter-intuitive given the fact that the signal $\mtt$ distribution is less peaked than $m_{W\gamma}$ (as shown in Fig.~\ref{mTcut}, some events always leak to lower values). However, we find that this effect is countered by the steeper dropoff of the $W\gamma$ background with $\mtt$. The dominance of $\Delta y$ over $\cos{\theta}_{CS}$ as the angular variable is less surprising based on the shape distributions presented in Fig.~\ref{Proton}. There, we saw that the signal $\Delta y$ distribution peaks at the center -- right at the background RV, whereas the signal $\cos{\theta}_{CS}$ peak is offset from the background RV.

The RV improved mass sensitivities are presented in Fig.~\ref{mass} for the different model scenarios. The left panels show the sensitivity improvement given the current LHC luminosity ($\sim 140\, \ifb$), while the extrapolation to the HL-LHC is shown in the panels to the right. As the combination of the RV cut on $\Delta y$ for events binned by $\mtt$ always delivered the best sensitivity, we only show those results.  In the following we describe each panel in more detail.

The top-left panel shows the sensitivity for scalar resonances of mass $m_X$ with current luminosity at the LHC. The black band marks the $2 \sigma$ exclusion using the $\mtt$ distribution alone whereas the gray band uses $\mtt$ along with the RV cut on $\Delta y$ as shown in the gray label. Values of the cross section above these bands are excluded. The width of the bands come from considering a systematic error between once and twice as large as the statistical error in each $\mtt$ bin. As we can see, applying the RV cut on $\Delta y$ improves the sensitivity by about $30 \%$. In other words, the RV-improved analysis is equivalent to increasing the current luminosity from $140\,\ifb$ up to about $250-300\, \ifb$ depending on the resonance mass region. In the same panel, the red line corresponds to the squirks production cross section. We assume that the squirks decay into $W\gamma$ $85\%$ of the time following \cite{Burdman:2008ek}, and only include two generations of squarks. The blue lines represent two benchmarks in the parameter space of the Scalar Triplet model: $\Lambda = 5 m_X$, $y_m = 0.1$ (solid) and $\Lambda = 10$ TeV, $y_m = 0.15$ (dot-dashed). The point where the signal lines (red and blue) intercept with the black and the gray bands correspond to the bounds on the resonance mass with and without implementing the RV angular cuts, respectively. Looking at the signal lines, we can see that using the RV cut on $\Delta y$ the LHC can increase the mass sensitivity by $80$ GeV for the squirk model and by 70 GeV (140 GeV) for the first (second) benchmark of the Scalar Triplet model.

These results depend on the parameters of the models we study. For example, the sensitivity bands (black) can change if the $W\gamma$ resonances are not narrow. This happens because a broad resonance will spread the events in the $\mtt$ distribution, washing out the signal. Similarly, the signal lines in Fig.~\ref{mass} can be modified in two ways. For the squirk model, the red line will move up (down) if the $BR$ of the squirks to $W\gamma$ increases (decreases). The slope of this line does not change because the production of squirks is fixed by the electroweak production of squarks as described in Section \ref{sec:models}. For the triplet models, the signal blue lines move up (down) as the coupling $y_m$ increases (decreases). Also, the slope of the signal line changes depending on the choice of $\Lambda$ as shown in the figure. The smaller the slope, the highest is the increase in sensitivity provided by the RV cut.

The top-right panel shows the high luminosity projections of the sensitivity for scalar resonances. As explained earlier, to calculate the bounds for HL-LHC we have rescaled the number of events in each distribution by the factor by which the luminosity increases.\footnote{This consideration is well justified in that our MC simulations were ran with enough iterations to suppress the statistical fluctuations at the sub-percent level.} We did this for the main background, the secondary background and the signal distributions. In the case of HL-LHC, applying the RV angular cuts the LHC can probe extra $150-300$ GeV of resonance masses, which is equivalent to running the LHC with a luminosity of $6-8\,{\rm ab}^{-1}$.

Moving to the middle-left and bottom-left panels, the green and purple lines correspond to the Vector Triplet $c_W$ and the Vector Triplet $c_h$ model, respectively. For the former we used the two benchmarks: $\Lambda = 5\,m_X$, $g_m = 0.03$ (solid) and $\Lambda = 10\,m_X$, $g_m = 0.06$ (dot-dashed), whereas for the latter we used the benchmarks $\Lambda = 5\, m_X$, $g_m = 0.005$ (solid) and $\Lambda = 10\, m_X$, $g_m = 0.02$ (dot-dashed). Again, the black bands provide the bounds on $m_X$ using the entire $\mtt$ distribution whereas the gray bands use $\mtt$ as well as the RV angular cuts. We can see how our analysis increases the mass sensitivity by $50 - 220$ GeV, the smallest increase corresponding to the Vector Triplet $c_h$ model.

\begin{figure}[t!]
	\includegraphics[width=1.0\linewidth]{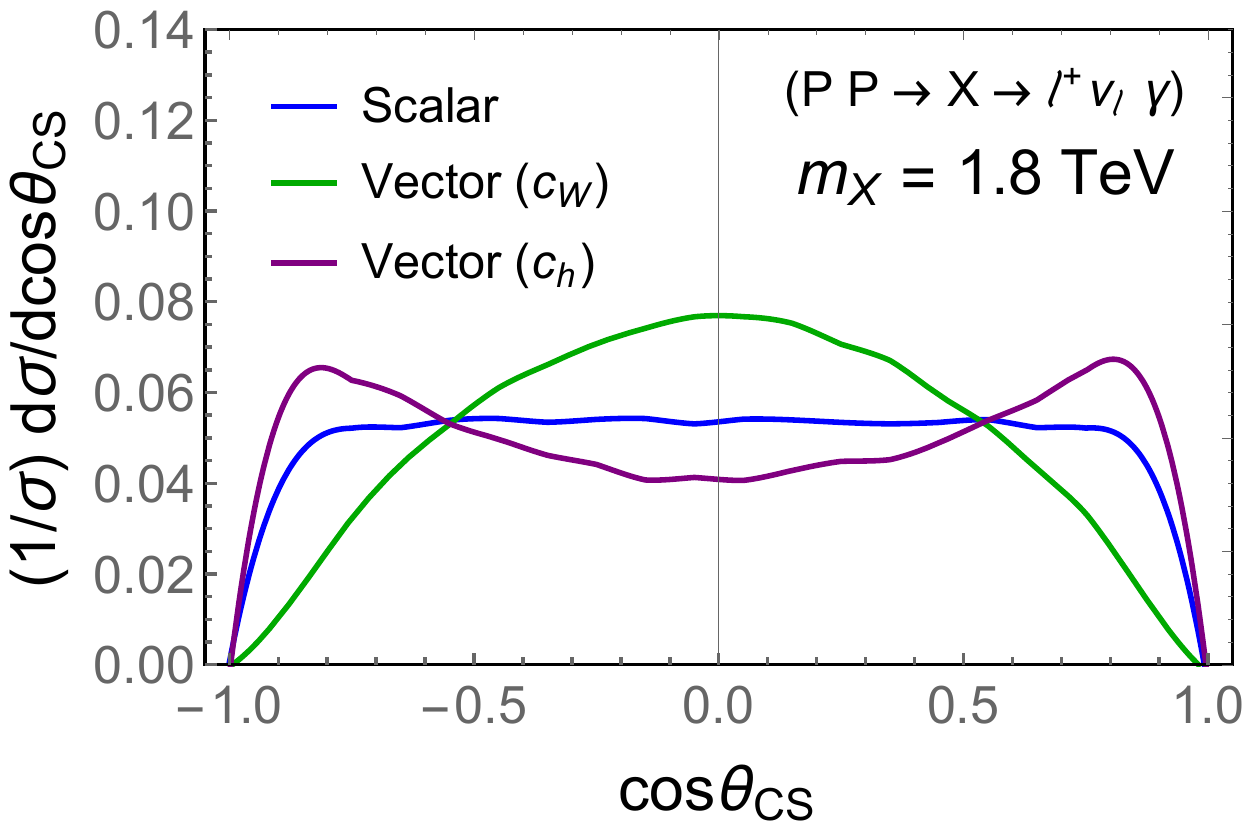}
\caption{\label{CS}{\small Signal shape of the CS angle for $W^+(\ell^+\nu)\gamma$ resonances. Heavy resonances show different shapes depending on the spin; scalar resonances decay isotropically whereas a spin one resonance creates different patterns depending on its couplings to gauge bosons.}}
\end{figure}

The middle- and bottom-right panels show the high luminosity projections of the sensitivity for the vector models. We can see how the mass sensitivity increases by $100 - 350$ GeV for vector resonances. This is equivalent to increasing the luminosity of the HL-LHC by a factor of 3, the most optimistic scenario corresponding to the Vector Triplet $c_W$ model.

Note that the height of the black and gray bands is different in each panel. This is because the signal efficiency is different for each model. For example, the Vector Model $c_W$ predicts a $\mtt$ distribution more peaky than the other models. Also, the Vector Model $c_h$ predicts a more spread $\mtt$ distribution so the efficiency of the RV cut is not as good as for the other models.

Our results focus on the $W\gamma$ decay of the scalar and vector resonances we have introduced. However, one can potentially look for these particles in dijet final states for they have couplings to quarks. Current searches at the LHC exclude dijet resonances with signal cross sections of 70 and 2 fb in the mass range between 1.2 and 6.5 TeV with $139 \ifb$ of data \cite{Aad:2019hjw}. Previous searches excluded signal cross sections above 0.1 pb and masses above 1 TeV (2 TeV) for $29.3 \ifb$ ($77.8 \ifb$) of data \cite{Aaboud:2018fzt,CMS:2018wxx,Sirunyan:2019pnb}. Most of the benchmarks we provided in Fig. \ref{mass} produce dijet cross sections that are well below current sensitivity. The only exception is the second benchmark ($\Lambda=10m_X$ and $g_m=0.02$) of the Vector Triplet $c_h$ model. For that parameter point, $BR(jj) \gg BR(W\gamma)$ because $g_m$ is not small enough to compensate for the large value of $\Lambda$. We verified that a resonance of about 700 GeV (close to the bounds in our Fig. \ref{mass}) should produce $\sim 1\sigma$ deviations in the current dijet background. With more data, this would imply correlated signals in dijets and $W\gamma$.

{\it The role of $\cs$ - } While we have focused on $\Delta y$ for the purposes of the RV cut, the $\cs$ variable is still a useful one. Should a $W\gamma$ resonance be found, the $\cs$ angle can also be used to discriminate the spin of NP signals \cite{Raj:2016aky,Capdevilla:2017doz} (Fig.~\ref{CS}). Scalar resonances decay isotropically, leading to a flat distribution in $\cs$ (except at the edges where the kinematic cuts of the final states objects reduces the number of events) while vector resonances populate either the edges or the central region depending on the couplings to the vector bosons (i.e. if one considers the Vector Triplet $c_W$ or $c_h$ model). The discriminatory power is best for heavy resonances, as they are less affected by acceptance cuts (compare the $\cs$ distribution for a scalar resonance in Fig.~\ref{CS} (1.8 TeV resonance) with the signal distribution in Fig.~\ref{Proton} (600 GeV scalar resonance)).

\section{Conclusions}\label{conclude}

In this paper, we showed that the Radiation Valley (RV) -- the remnant of the Radiation Amplitude Zero (RAZ) at a hadron collider -- can play an important role in the discovery of exotic resonances that decay to $W(\ell \nu)\gamma$. While the RAZ is encoded in several different kinematic variables, we found that the best proxy for it is the difference in rapidity between the lepton and the photon in the case of resonance searches. Adapting the ATLAS $W(\ell\nu) \gamma$ search~\cite{Aad:2014fha} to include a RV cut, we explored the improvement in resonances for four different straw-man resonance scenarios: scalar squirks, a phenomenological scalar triplet model, and two different phenomenological vector resonance models. Optimizing the RV cut for the resonance mass and type, we found an increase of order 70 - 220 GeV in the resonance mass reach assuming $140\, \rm{fb}^{-1}$ of integrated luminosity (an estimate of the current run II dataset).  Extrapolating to the full HL-LHC dataset, we project the increase in reach to be of order 100 - 350 GeV. The increase in mass reach that we find is equivalent to increasing the luminosity by a factor between 2 and 3, depending on the mass region, model assumptions, and luminosity. We focused exclusively on leptonic $W$ to avoid contamination from fake backgrounds, such as $\gamma + \text{jet}$, which do not share the RAZ features of the irreducible background.  Jet substructure and, more recently, jet image based searches have shown discriminating power between hadronic $W$ and QCD~\cite{Aaboud:2018fgi,Fraser:2018ieu,Chen:2019uar,deOliveira:2015xxd}, therefore it would be interesting to combine those techniques with RV cuts in $W\gamma$ resonance searches with hadronic $W$. Finally, even though we focused on the effects of resonances on the RAZ, exploring non-resonant effects is another interesting venue. In such a case, angular asymmetries can play an important role in signal-to-background discrimination.

\section*{Acknowledgments}

We thank Christopher Kolda, Antonio Delgado, Michele Grossi, Jakob Novak, and Wai-Yee Keung for useful discussions. We also thank Ketevi Assamagan and Alessandro Tricoli for discussions at the Brookhaven National Laboratory. RC thanks Tao Han for his invitation to PITT PACC and for the fruitful discussions. The work of AM was partially supported by the National Science Foundation under Grants No. PHY-1820860. The work of RC was supported in part by the Canada Research Chair program and by the Perimeter Institute for Theoretical Physics (PI). Research at PI is supported by the Government of Canada through Industry Canada and by the Province of Ontario through the Ministry of Economics Development and innovation. The work of RH was supported by the DOE under contract number DE-SC0007859 and Fermilab, operated by Fermi Research Alliance, LLC under contract number DE-AC02-07CH11359 with the United States Department of Energy.

\begin{center}
\rule{5cm}{0.6pt}
\end{center}

\bibliographystyle{unsrt}
\bibliography{xampl}

\end{document}